\documentclass[aip]{revtex4-1}
\usepackage{color,graphicx,siunitx,textcomp,amssymb,amsmath,dcolumn,hyperref,amsthm}
\newcommand{\ket}[1]{|#1\rangle}
\begin{document}
\title{Kondo effect and spin--orbit coupling in graphene quantum dots}
\author{A.~Kurzmann}
\email{annikak@phys.ethz.ch}
\affiliation{Solid State Physics Laboratory, ETH Zürich, CH-8093 Zürich, Switzerland}
\author{Y.~Kleeorin}
\affiliation{Center for the Physics of Evolving Systems, Biochemistry and Molecular Biology,
University of Chicago , Chicago, IL, 60637, USA}
\author{C.~Tong}
\author{R.~Garreis}
\affiliation{Solid State Physics Laboratory, ETH Zürich, CH-8093 Zürich, Switzerland}
\author{A.~Knothe}
\affiliation{National Graphene Institute, University of Manchester, Manchester M13 9PL, United Kingdom}
\author{M.~Eich}
\author{C.~Mittag}
\author{C.~Gold}
\author{F.K.~de Vries}
\affiliation{Solid State Physics Laboratory, ETH Zürich, CH-8093 Zürich, Switzerland}
\author{K.~Watanabe}
\affiliation{Research Center for Functional Materials, National Institute for Materials Science, 1-1 Namiki, Tsukuba 305-0044, Japan}
\author{T.~Taniguchi}
\affiliation{International Center for Materials Nanoarchitectonics, National Institute for Materials Science,  1-1 Namiki, Tsukuba 305-0044, Japan}
\author{V.~Fal'ko}
\affiliation{National Graphene Institute, University of Manchester, Manchester M13 9PL, United Kingdom}
\author{Y.~Meir}
\affiliation{Department of Physics, Ben-Gurion University of the Negev, Beer-Sheva, 84105, Israel}
\author{T.~Ihn}
\author{K.~Ensslin}
\affiliation{Solid State Physics Laboratory, ETH Zürich, CH-8093 Zürich, Switzerland}

\date{\today}

\begin{abstract}
\textbf{The Kondo effect is a cornerstone in the study of strongly correlated fermions. The coherent exchange coupling of conduction electrons to local magnetic moments gives rise to a Kondo cloud that screens the impurity spin\cite{kondo1964resistance}. Whereas complete Kondo screening has been explored widely, realizations of the underscreened scenario — where only some of several Kondo channels participate in the screening — remain rare\cite{goldhaber1998kondo,cronenwett1998tunable}. Here we report the observation of fully screened and underscreened Kondo effects in quantum dots in bilayer graphene. More generally, we introduce a unique platform for studying Kondo physics. In contrast to carbon nanotubes\cite{nygaard2000kondo,paaske2006non,makarovski20072}, whose curved surfaces give rise to strong spin--orbit coupling\cite{kuemmeth2008coupling,steele2013large} breaking the SU(4) symmetry of the electronic states relevant for the Kondo effect\cite{jespersen2011gate,cleuziou2013interplay}, we study a nominally flat carbon material with small spin--orbit coupling. Moreover, the unusual two-electron triplet ground state\cite{kurzmann2019excited} in bilayer graphene dots provides a route to exploring the underscreened spin-1 Kondo effect.}
\end{abstract}
\maketitle
\section*{Introduction}
Electronic conduction at low temperatures can be affected by a small amount of magnetic impurities, a phenomenon known as the Kondo effect \cite{kondo1964resistance}. The spin of a localized electron coherently couples to the spins of delocalized electrons in the host material, resulting in a net spin of zero and the formation of the Kondo screening cloud \cite{wilson1975renormalization}. Due to phase-space constraints, mainly electrons near the Fermi surface are affected, leading to the characteristic signature of the Kondo effect: a narrow resonance at the Fermi energy. The effect has been observed in a variety of materials, including graphene, with impurities acting as the host for the localized spin. Kondo temperatures in graphene were shown to reach up to 90 K \cite{chen2011tunable}.

The Kondo effect was discovered experimentally in semiconductor quantum dots in 1998 \cite{goldhaber1998kondo,cronenwett1998tunable}, confirming theoretical predictions\cite{Ng88,Glazman88,Meir93,Wingreen94,Izumida98}. Quantum dots with their net spins act as magnetic artificial atoms, and nearby leads take the role of the surrounding Fermi sea. The high tunablity of quantum dots is an important aspect of studies of the Kondo effect \cite{van2000kondo}. Measurements on electrostatically defined quantum dots in GaAs enabled the observation of the unitary limit of the Kondo effect \cite{van2000kondo}. In studies of quantum dots in carbon nanotubes the singlet triplet \cite{paaske2006non} and the SU(2) and SU(4) Kondo effect were explored \cite{makarovski20072}. Moreover, spin--orbit interaction in carbon nanotubes \cite{kuemmeth2008coupling} were found to change significantly the low-energy Kondo physics, which often complicates studies of strongly correlated effects.

Improvements in the fabrication of nanostructures\cite{eich2018spin,banszerus2018gate,kurzmann2019excited} in 2D materials pave the way to reveal Kondo physics in quantum dots electrostatically defined in a flat bilayer graphene sheet with small spin--orbit coupling. In addition an unusual two-hole triplet ground state\cite{kurzmann2019excited}, and an exceptional tunability of tunnel rates, dot size and valley magnetic moment\cite{tong2020tunable}are present. The interplay between a small spin--orbit coupling and the Kondo effect is further known to lead to an underscreened spin-1 Kondo effect, in which the net spin of the conduction electrons only partially compensates the localized spin.  This effect has been observed in mechanically controlled individual cobalt complexes\cite{parks2010mechanical}, but not in quantum-dot systems, which have the advantage of being electrically tunable. 

\section*{Kondo effect in bilayer graphene}

\begin{figure*}
	\includegraphics[scale=0.9]{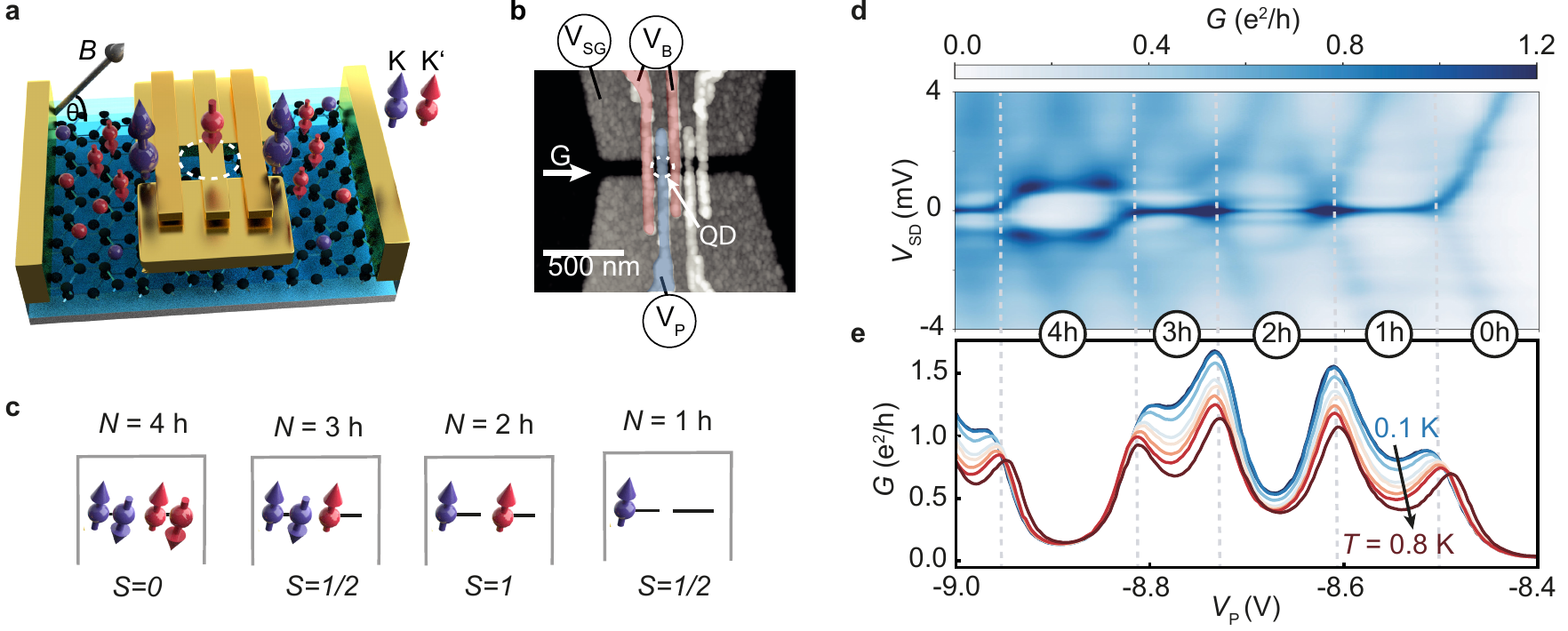}
	\caption{Kondo effect in bilayer graphene quantum dot devices. \textbf{a}, Schematic of the quantum dot device in bilayer graphene. A magnetic field is applied at an angle $\Theta$ relative to the graphene flake. \textbf{b}, Atomic force micrograph of the gate structure on top of the bilayer graphene stack. Three finger gates are used to form the quantum dot (dashed white circle), with two tunable barriers (red) and a plunger gate (blue). \textbf{c}, Level-filling scheme of the QD with the number of charge carriers $N$. \textbf{d}, Finite-bias measurement of a quantum dot that is strongly coupled to the leads. The grey dashed lines indicate plunger-gate voltages where additional holes enter the dot. A finite conductance is observed in the Coulomb blockaded regions (between dashed lines) at $V_\text{SD}=0\,\text{mV}$, when the dot is charged with $N=0$ to $N=4$ holes. \textbf{e}, Typical temperature dependence of Kondo resonances at a low bias voltage ($V_{\text{SD}}=\SI{2}{\micro eV}$).}
	\label{fig1}
\end{figure*}

Our gate-defined quantum dots in bilayer graphene (shown in Fig.~\ref{fig1}a and b) are investigated through two-terminal AC measurements (see methods for details). The low-temperature differential conductance $G$ of a strongly coupled dot is presented in Fig.~\ref{fig1}d as a function of DC source--drain bias $V_\textrm{SD}$ and plunger gate voltage $V_\textrm{P}$. The regions of low conductance (white) are caused by Coulomb blockade, whereas the lines of higher conductance (light blue) are edges of the Coulomb blockade diamonds. Within each diamond, the quantum dot hosts a fixed integer number $N$ of holes. When filled with many holes, a four-hole shell-filling periodicity emerges\cite{eich2018spin,garreis2021shellfilling,knothe2020quartet}, which reflects the four-fold degeneracy of the graphene spectrum: one factor of two from the spin and one from the isospin (K, K') that stems from the valley degree of freedom.

Most importantly, the differential conductance exhibits lines of high conductance centred at $V_{\mathrm{SD}}= 0$ within the $N=1$, $2$, and $3$ Coulomb diamonds, which is a signature of Kondo-assisted tunneling through the quantum dot. The zero-bias resonance is absent for $N=4$, where all spin and valley states pair up (see Fig.~\ref{fig1}c). For odd filling ($N=1,3$), the QD is charged with spin $S=1/2$, leading to a mixing of the spin and valley Kondo effect. For half-filling ($N=2$), a weaker Kondo resonance is observed. The QD is charged with two holes with the same spin and the spin--1 Kondo effect is expected. The presence of the Kondo effect is confirmed by the temperature dependence of the zero-bias conductance as a function of gate voltage (Fig.\ref{fig1}e). The individual traces taken at different temperatures show an increasing conductance with decreasing temperature for $N=1,2,3$.

\section*{Spin--orbit splitting in Kondo measurements for $N=1$}

 \begin{figure*}
	\includegraphics[scale=0.9]{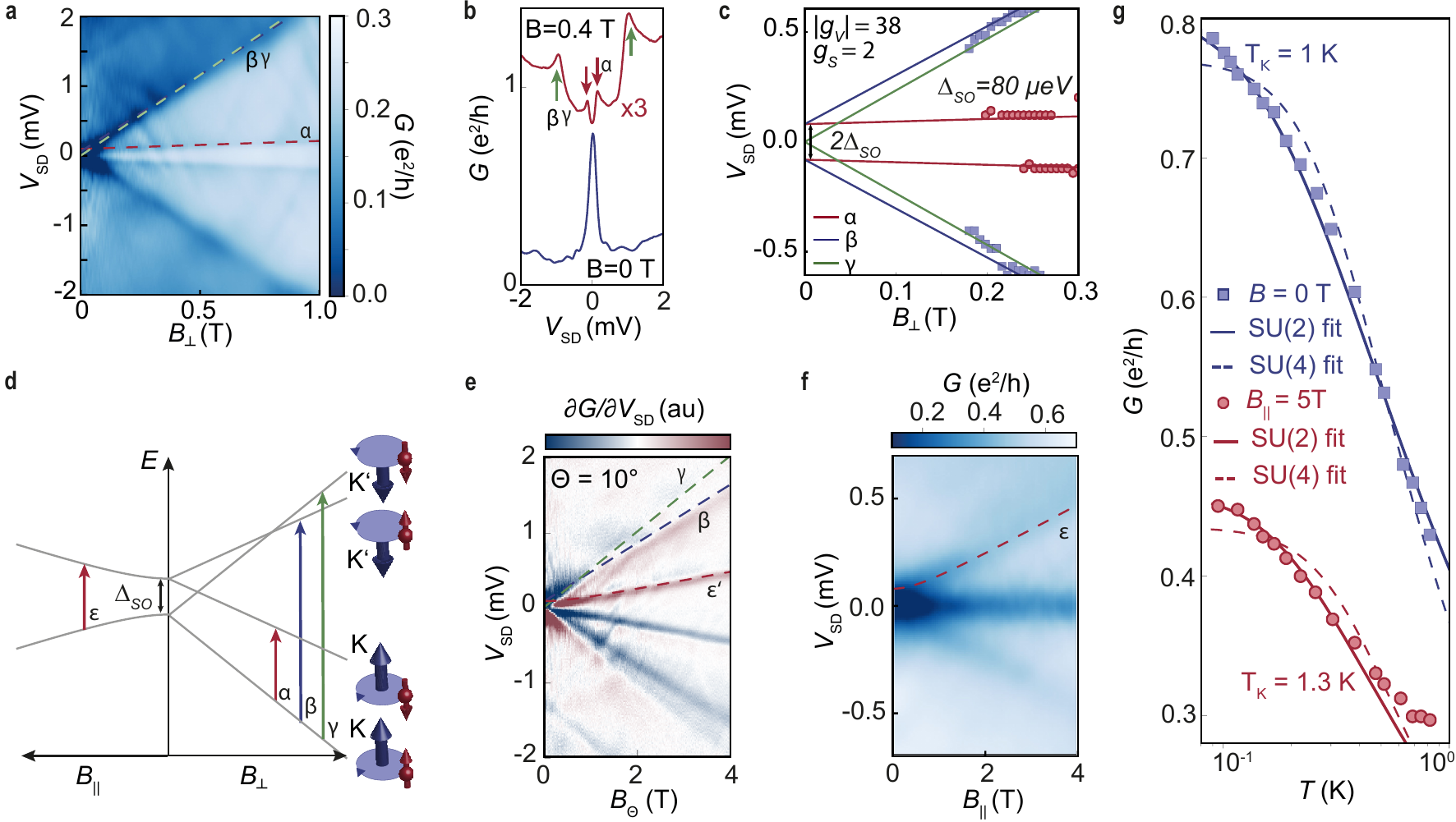}
	\caption{Interplay between the spin--orbit coupling and the Kondo effect for $N=1$. \textbf{a}, Experimentally observed splitting of the Kondo resonance into four resonances in a perpendicular magnetic field. The measurements are fully described by the energy levels in \textbf{d} with a valley $g$-factor of 38, an electronic g-factor of 2 and a spin--orbit splitting of \SI{80}{\micro eV}. \textbf{b}, Line cuts in \textbf{a} at $B=0$ and \SI{0.4}{T}.  \textbf{c} Evaluation of the conductance maxima in \textbf{a} using a peak finding algorithm. \textbf{d}, Energy levels in bilayer graphene quantum dots in parallel and perpendicular magnetic fields, with a zero magnetic field splitting due to the spin--orbit coupling $\Delta_{\mathrm{SO}}$. The alignment of the magnetic moments is shown with four small icons, where the red arrow represents the spin magnetic moment and the blue arrow the valley magnetic moment. A parallel alignment between the spin and valley magnetic moments is preferred.  \textbf{e}, Derivative of measurements of the Kondo resonance with an angle of 10° between the magnetic field and the sample plane showing all energy transitions. \textbf{f}, Splitting of the Kondo resonance in a parallel magnetic field. The Kondo resonance splits into three resonances: two from the spin splitting and one from the valley Kondo resonance that is not influenced by a parallel magnetic field.  \textbf{g}, Temperature dependence of the maximum conductance of the Kondo resonance, showing a lifting of the SU(4) symmetry, due to the spin--orbit coupling. For zero magnetic field the SU(2) Kondo model with a Kondo temperature of 1 K describes the experimental data well.}
	\label{fig2}
\end{figure*}

In Fig.~\ref{fig2}a, we show how for $N=1$ the single Kondo peak at $B=0$ splits into four peaks at finite magnetic field, when measured in the centre of the $N=1$ diamond at constant $V_\textrm{P}$. This is made visible in Fig.~\ref{fig2}b by measuring line cuts at $B=0$ and \SI{0.4}{T}. The observed split peaks are labeled $\alpha$ (red arrows), and $\beta,\gamma$ (green arrows) in Figs.~\ref{fig2}a,b.  We use a peak-finder algorithm to identify the conductance peaks $\alpha$ and $\beta$ from Fig~\ref{fig2}a and plot them as blue and red data points in Fig.~\ref{fig2}c, respectively. The slope of the conductance resonance $\alpha$ is described by the spin $g$-factor of two. Extrapolating the conductance peak belonging to this resonance with a straight line to $B=0$ (red dashed lines in Fig.~\ref{fig2}a) gives a zero-field splitting $\Delta_{SO}=$\SI{80}{\micro eV}, which we interpret as a spin--orbit splitting; note that this splitting is not directly resolved at $B=0$ in Fig.~\ref{fig2}(a,b). The observed splitting is of the same order of magnitude as the spin--orbit gap in previous measurements in quantum point contacts in bilayer graphene \cite{banszerus2020observation} and more than a factor of three smaller than in quantum dots in carbon nanotubes\cite{nygaard2000kondo,laird2015quantum}. The discrepancy is due to the different origin of the spin--orbit coupling: in nanotubes it originates from the curvature of the tube, while the spin--orbit coupling in graphene is of Kane-Mele type\cite{kane2005quantum}. The theoretical predicted Kane-Mele gap induced by phonons is of the order of 0.1 meV\cite{ochoa2012spin}, in good agreement with our measurements.

 Fig.~\ref{fig2}d shows the one-hole energy-level spectrum. The respective alignment of the magnetic moments is shown with four small icons, where the red arrow represents the spin magnetic moment and the blue arrow the valley magnetic moment. In the presence of  spin--orbit coupling, it is composed of two Kramer pairs split by the spin--orbit gap. The ground state exhibits a preferred parallel alignment of the valley and the spin magnetic moments. Applying a perpendicular magnetic field separates each energy doublet into two states with slopes proportional to either $g_\mathrm{v}\pm g_\mathrm{s}$ or $-g_\mathrm{v}\pm g_\mathrm{s}$, with valley and spin g-factors $g_\mathrm{v}$ and $g_\mathrm{s}$, respectively. This one-hole level scheme results in the three magnetic field-dependent excitations $\alpha,\beta$ and $\gamma$ (see Fig.~\ref{fig2}).

The conductance peak shown in blue in Fig.~\ref{fig2}c includes the two expected excitations $\beta$ and $\gamma$, which are difficult to separate experimentally (blue and green lines in Fig.~\ref{fig2}c). This is due to the strong valley splitting with $\mid g_\mathrm{v}\mid=38$ and the finite $V_{SD}$ bias window, which limits the observation of these transitions to small $B_\perp$. However, the valley splitting depends only on the perpendicular component of the magnetic field and is shifted to lower $V_\textrm{SD}$ when the sample is rotated by \SI{80}{\degree}. The spin splitting remains constant upon rotation as it depends on the total magnetic field. Therefore both excitations $\beta$ and $\gamma$ can be observed separately at higher magnetic fields above 2 T as shown in Fig.~\ref{fig2}e. At large enough field the spins now align with the almost parallel magnetic field leading to a transition $\epsilon'$ similar to $\epsilon$. This measurement confirms the presence of the energy level spectrum shown in Fig.~\ref{fig2}d.

In a magnetic field applied parallel to the sample (Fig.~\ref{fig2}f), the spin degeneracy of the zero-field spectrum is lifted while the valley states are not affected. For $N=1$ the Kondo resonance splits into three peaks (Fig.~\ref{fig2}f). The outer two resonances (marked with red dashed lines) split off as described by $g_\textbf{s}=2$. Within our model, the resonance at $V_\textrm{SD}=0$ and  finite parallel field is a pure valley Kondo resonance with fluctuations of the state in the dot between $K\uparrow$ and $K'\uparrow$. 

In the absence of spin--orbit interaction an SU(4) Kondo model will describe the zero-bias resonance for $N=1$ at zero magnetic field, with a characteristic temperature dependence of the Kondo peak conductance\cite{keller2013observation}. Deviations from this prediction due to the presence of the spin--orbit gap can therefore be studied via the temperature dependence of the maximum conductance of the Kondo resonance at $V_{\mathrm{SD}}=0$ and $B=0$. The corresponding data (blue data points in Fig.~\ref{fig2}g) can be fitted with a model for the temperature dependence of the SU(2) Kondo effect with spin $S=1/2$, given approximately by  \cite{goldhaber1998kondob} 
\begin{equation}
G(T)=G_0\left(\frac{T'^2_\textrm{K}}{T^2+T'^2_\textrm{K}}\right)^s
\end{equation}
with $T'_\textrm{K}=T_\textrm{K}/(2^{1/s}-1)^{1/n}$, where $s=0.22$ and $n=2$ for a spin-1/2 system, $G_0$ is the amplitude of the peak, $T_\textrm{K}$ the Kondo temperature and $T$ the electron temperature. We compare this fit (blue solid line in Fig.~\ref{fig2}g) with the corresponding fitted temperature dependence of the SU(4) Kondo model with $s=0.2$ and $n=3$  (ref.~\cite{keller2013observation}; blue dashed line). The data agree better with the SU(2) Kondo model. Numerical renormalization calculations (see SI) confirm that a small spin--orbit energy splitting will lead to a better fit to the peak by the SU(2) form. Yet it is not enough to split the zero-bias peak (calculations shown in the SI). (Another reason for the apparent transition from SU(4) to SU(2) are different tunnel rates of different channels \cite{kleeorin2017abrupt}.) In summary, the measurements of N = 1 h can be completely described by a SU(2) Kondo effect due to spin--orbit splitting of the 4-fold degeneracy.

\section*{Underscreened Spin 1 Kondo effect for $N=2$}

\begin{figure*}
	\includegraphics[scale=0.9]{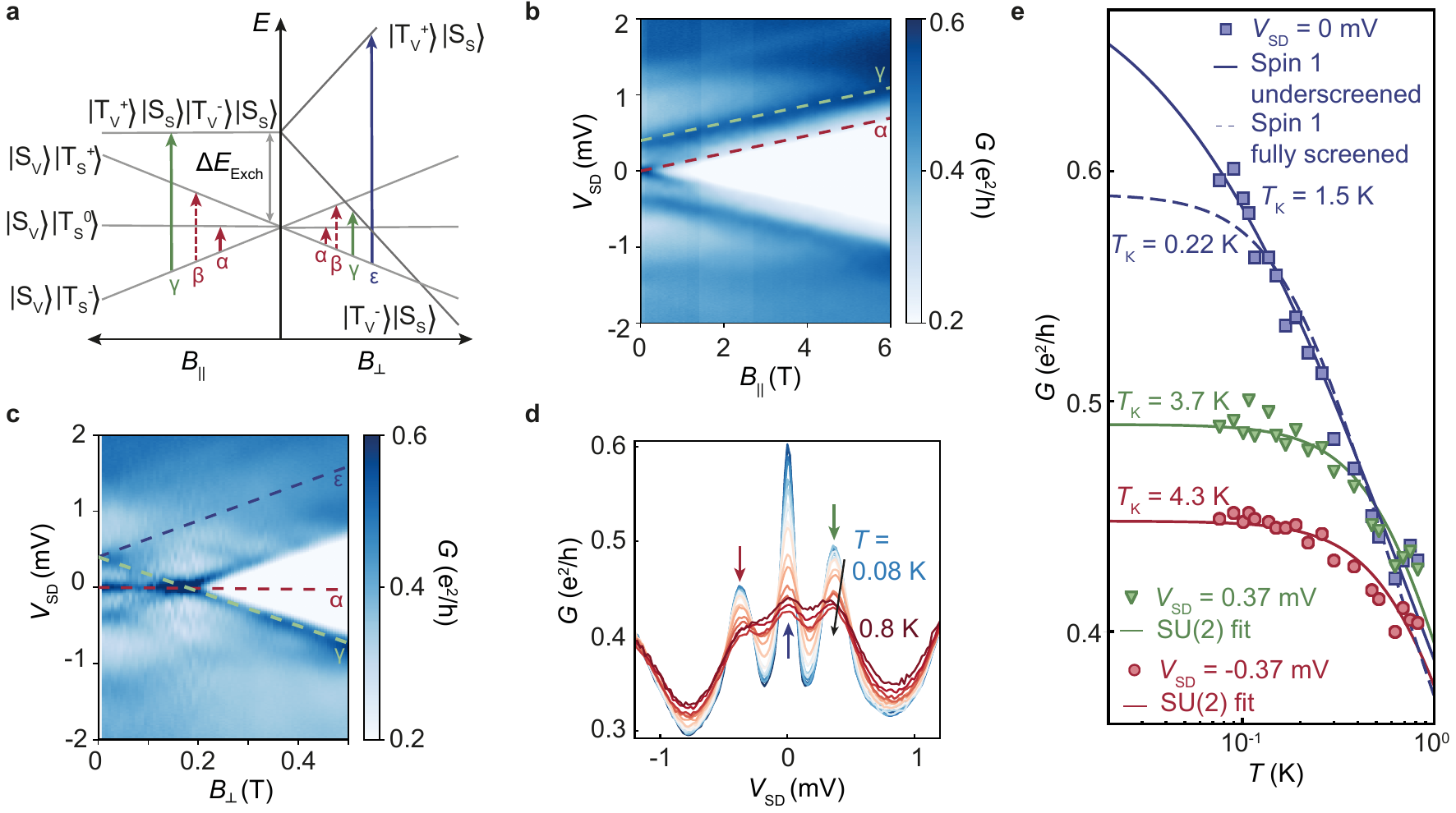}
	\caption{Triplet Kondo effect for $N=2\,\text{h}$. \textbf{a}, Expected level splittings in magnetic field applieds parallel and perpendicular to the sample with exchange energy $\Delta E_{\mathrm{Exch}}$. \textbf{b}, Kondo resonance in parallel magnetic field. A splitting of the Kondo resonance with the spin $g$-factor of two and an energy increase of the resonances at finite bias is observed. \textbf{c}, Splitting of the Kondo resonance in perpendicular magnetic field. The resonances at finite bias split with a valley $g$-factor of 40, while the Kondo resonance is not affected by the magnetic field up to 0.2 T.  \textbf{d}, Conductance around $V_{\textrm{SD}}=0$ at $B=0$ for different temperatures. A Kondo resonance at $V_{\textrm{SD}}=0$ and two finite-bias resonances at $V_{\textrm{SD}}=\pm0.37\,\text{mV}$ are observed. \textbf{e}, Temperature dependence of the Kondo resonance at $V_{\mathrm{SD}}=0$ and the finit-bias resonances at $V_{\mathrm{SD}}=\pm0.37\,\text{meV}$. The temperature dependence can be described by a spin-1 Kondo model with a Kondo temperature of $\SI{1.5}{K}$. The finite-bias resonances are fitted with a SU(2) Kondo model and Kondo temperatures of about $\SI{4}{K}$.}
	\label{fig3}
\end{figure*}

Figure~\ref{fig3}a shows the magnetic field dependent energy spectrum for $N=2$\cite{kurzmann2019excited}. The six two-hole states  can be constructed from linear combinations of the four degenerate spin and valley single-particle states. Exchange  interaction splits these states into a spin-triplet ground state (labeled $\ket{S_V}\ket{T_S^-}$, $\ket{S_V}\ket{T_S^0}$ and $\ket{S_V}\ket{T_S^+}$ in Fig.~\ref{fig3}a) that is three-fold degenerate at zero magnetic field, a spin-singlet state with a two-fold valley degeneracy (labeled $\ket{T_V^-}\ket{S_S}$, $\ket{T_V^+}\ket{S_S}$) at zero magnetic field, and a single spin-singlet valley-triplet state at the highest  energy (not shown)  \cite{kurzmann2019excited}. Applying a parallel magnetic field, the triplet ground state  splits into its three spin components, while all the other excited states remain unaffected. Applying a perpendicular magnetic field splits the valley-triplet spin-singlet excited state, leading to a strong energy reduction of $\ket{T_V^-}\ket{S_S}$ with increasing magnetic field and a strong energy increase of $\ket{T_V^+}\ket{S_S}$. The $\ket{S_V}\ket{T_S^-}$ sets the origin for the co-tunneling excitations in parallel magnetic field and for small perpendicular fields. The excitations (shown as vertical arrow and labeled $\alpha$, $\beta$, $\gamma$, and $\epsilon$) belong to the transitions from the ground state to the excited states shown in Fig.~\ref{fig3}a.

Figures~\ref{fig3}b shows the splitting of the Kondo resonances as a function of $B_\parallel$. The resonances $\alpha$ and $\gamma$ introduced in Fig.~\ref{fig3}a are observed and marked with red and green dashed lines. Slope and offset are given by the spin Zeeman effect ($g_\mathrm{s}=2$) and the exchange energy $\Delta E_{\mathrm{Exch}}=0.37$ meV. Resonance $\alpha$ appears as a splitting of the $B=0$ Kondo resonance,  whereas resonance $\gamma$ is seen as a cotunneling resonance already at $B=0$ but finite $V_\textrm{SD}$ (see also Fig.~\ref{fig1}c). Because the excited state energy is independent of $B_\parallel$, this resonance runs in parallel to the split Kondo resonance. The expected transition labeled $\beta$ is not seen experimentally in Fig.~\ref{fig3}b and c as it would require spin flips of both holes. 

Corresponding data for $B_\perp$ are depicted in Fig.~\ref{fig3}c. The Kondo resonance $\alpha$ (red dashed line) is observed at $B_\perp<B_\perp^\star=\SI{0.2}T$, but its Zeeman splitting remains unobservable at these low fields. The cotunneling transition $\gamma$ (green dashed line), split-off from $V_\textrm{SD}$ by the exchange energy at $B_\perp=0$, moves down linearly with increasing $B_\perp$ according to the valley Zeeman effect, hitting $V_\textrm{SD}=0$ at $B_\perp^\star$, as expected from Fig.~\ref{fig3}a. The extracted $\mid g_\mathrm{v}\mid=38$ is in excellent agreement with the valley splitting determined for $N=1$. At $B_\perp^\star$, the excitations $\alpha$ and $\gamma$ converge, resulting in an enhanced zero-bias peak. The state $\ket{T_V^-}\ket{S_S}$ is the system ground state for $B_\perp>B_\perp^\star$ the transition to the $\ket{S_V}\ket{T_S^-}$ state is the experimentally dominating cotunneling transition involving a simultaneous valley and spin flip.
 
Figure~\ref{fig3}d  shows the differential conductance at $B=0$ for temperatures between \SI{0.08}{K} (blue) and \SI{0.8}{K} (red). The finite-bias resonances at $V_{\mathrm{SD}}=\pm 0.37\,mV$ are marked by red and green arrows and the Kondo resonance by a blue arrow. We study the temperature-dependent conductance of the three resonances in detail in Fig.~\ref{fig3}e, where the blue data points belong to the Kondo resonance, while the red and green data points belong to the finite-bias resonances. 

Due to the spin triplet ground state in bilayer graphene quantum dots, we expect a spin-1 Kondo effect. Localized spin-1 magnetic impurities connected to two conduction channels each with $s=1/2$ screening capacity can be screened by two (fully screened), one (underscreened) or none (non-Kondo regime) of the channels. In Fig.  3e we show fits to  the data points using the underscreened  \cite{parks2010mechanical} (blue solid line) and fully screened (blue dashed line) spin-1 Kondo models. The underscreened Kondo model is in better agreement with the data, in particular at the lowest temperatures (for details of the fitting parameters see the NRG calculation results in the SI). The fully screened spin-1 model required exact symmetry of the two channels. Thus any splitting, e.g.,  due to spin--orbit interaction, will lead to a difference between the Kondo temperatures of each conduction channel, and to a two-stage Kondo effect\cite{pustilnik2001kondo}.  The data agree with such a scenario where the temperature $T$ lies around the first-stage Kondo temperature, such that only one channel participates in the screening.

\section*{Conclusion}
We have studied the breaking of the SU(4) symmetry of the Kondo effect for $N=1$ due to spin--orbit coupling with a magnitude of $\SI{80}{\micro eV}$. This spin--orbit coupling strength is in agreement with the theoretical estimate of the Kane-Mele gap induced by phonons\cite{ochoa2012spin}. Furthermore, the spin triplet ground state for $N=2$ allows us to study the dependence of the spin-1 Kondo resonance on magnetic field as well as temperature together with a small spin--orbit interaction in the QD. This spin--orbit interaction in bilayer graphene can lead to an underscreening of the spin in the quantum dot. The spin-triplet ground state, together with the wide range of tunability, makes quantum dots in bilayer graphene an unique model system for studying the underscreened spin-1 Kondo effect as well as the interplay between spin--orbit coupling and the Kondo effect in graphene nanostructures without a curved surface.

\section*{Methods}
The investigated bilayer graphene flake is embedded between two hBN flakes (20 nm and 39 nm thick) and stacked on top of a graphite back gate using the dry transfer technique. Source and drain contacts are fabricated by etch contacting \cite{wang2013one}. The gate structure fabricated on top of the stack is shown in the atomic force microscopy (AFM) image in Fig.~\ref{fig1}b. The split gates (shown in grey) can be used to form a 800 nm long and 100 nm wide channel in the bilayer graphene flake. As a second layer of gates, five finger gates (20 nm wide) are  deposited on top of an $\text{Al}_2\text{O}_3$ layer. Back and top gates can be used to (i) open a band gap below the gates and (ii) tune the Fermi energy into the band gap, rendering these regions insulating. An n-type channel is formed between the split gates by applying a positive voltage to the graphite back gate ($V_{\text{BG}}=3.7\,\text{V}$) and a negative voltage to the split gates ($V_{\text{SG}}=-2.92\,\text{V})$. An in-plane source-drain bias voltage $V_{SD}$ is applied to the channel using the pair of Ohmic contacts.

A fully tunable quantum dot can be formed in the channel using the three finger gates (colored red and blue) in Fig. 1b. The finger gate on top of the quantum dot (colored blue) is used to form an n-type quantum dot and controls the number of charge carriers in the quantum dot. The outer two finger  gates (colored red) are used to tune the tunnel coupling over wide ranges by gate voltages (see Ref.~\cite{tong2020tunable} for details). We can deplete the dot down to the last electron as seen from Coulomb blockade resonances. 

The electrical properties are investigated through two-terminal AC measurements, in which a variable DC voltage $V_\textrm{SD}$  and a AC component $V_{\text{AC}}=0.020\,\text{mV}$ are applied between source and drain contacts, where the differential conductance $G=\partial I/\partial V_\textrm{SD}$ is measured by standard lock-in techniques. The hole occupancy of the dot is controlled by the centre finger gate (shown in blue in Fig.~\ref{fig1}b) by a voltage $V_\textrm{P}$. We measure the device in a $^3$He/$^4$He dilution refrigerator with a base temperature of \SI{80}{mK}, fitted with a rotatable sample stick for out-of-plane rotations of the sample in magnetic fields of up to \SI{8}{T}.

\bibliographystyle{naturemag.bst}

\section*{Acknowledgments}
We thank Peter Märki, Thomas Bähler as well as the staff of the ETH cleanroom facility FIRST for their technical support. We also acknowledge financial support by the European Graphene Flagship. Growth of hexagonal boron nitride crystals was supported by the Elemental Strategy Initiative conducted by the MEXT, Japan, Grant Number JPMXP0112101001, JSPS KAKENHI Grant Number JP20H00354 and the CREST(JPMJCR15F3), JST. We acknowledge funding from the European Union’s Horizon 2020 research and innovation program under the Marie Skłodowska-Curie Grant Agreement No. 766025. YM acknowledges the support of the Israel Science Foundation (grant no.359/20). 

\subsection*{Authors contributions}
A.Ku. performed the experiments and fabricated the sample with help of M.E., C.T. and R.G.. A.Ku. analyzed the data with the assistance of C.G., M.E., C.T., R.G., C.M. and F.V.. K.W. and T.T. synthesized the hBN crystals. Y.K. and Y.M. developed the theoretical understanding and Y.K. performed NRG calculations. A.Kn. and V.F. developed the theoretical understanding of the spin--orbit coupling. Y.M., T.I. and K.E. supervised the project. All authors discussed the results.

\subsection*{Data availability}
The data supporting the findings of this study will be made available via the ETH Research Collection.

\subsection*{Code availability}
The computer code and the numerically calculated datasets generated during and/or analysed during the current study are available from the corresponding authors upon reasonable request.

\subsection*{Competing interests}
The authors declare no competing interests. 

%Here you should list the contents of your Supplementary Materials -- below is an example. 
%You should include a list of Supplementary figures, Tables, and any references that appear only in the SM. 
%Note that the reference numbering continues from the main text to the SM.
% In the example below, Refs. 4-10 were cited only in the SM.     
\section*{Supplementary materials}	
Details on NRG calculations\\
Figs. S1 and S2\\
References \cite{toth2008density,legeza2008manual,parks2010mechanical,blesio2019fully}

\section*{Supplementary Information: Kondo effect and spin-orbit coupling in graphene quantum dots}
\section*{NRG calculation}
We model the quantum dot in bilayer graphene system by a degenerate state with two quantum numbers $\sigma=\pm 1/2$ spin denoted $ \uparrow, \downarrow$ and $v=\pm 1/2$ valley denoted  $k, k'$ 
\begin{equation}
    H_{dot}=\sum_{\sigma v} (\epsilon_d +\frac{1}{2} \Delta_{so} v\sigma) n_{\sigma v} + U\sum_{\{\sigma v\}\ne\{\sigma' v'\}} n_{\sigma v}n_{\sigma' v'}+J\vec{S}_{k}\cdot \vec{S}_{k'} 
\end{equation}
where $n_{\sigma v}=d^\dagger_{\sigma v}d_{\sigma v}$ is the occupation number operator of valley $v$ with spin $\sigma$, $\Delta_{so}$ is the energy gap between the two spin-orbit split Kramer pairs (each pair characterized by a different value of the product $v\sigma$), $U$ is the Coulomb repulsion between electrons in different states , and $J$ is the magnetic interaction between electrons in the two valleys, that ensures a valley-singlet two-particle ground state, as seen experimentally. The spin vector for valley $v$ is defined as $\vec{S}_v=\frac{1}{2}\sum_{\sigma,\sigma'}d^\dagger_{v \sigma}\vec{\sigma}_{\sigma \sigma'}d_{v \sigma'}$, where $\vec{\sigma}$  are the Pauli matrices. 
The system is connected to leads on both sides $s=L/R$ by the tunneling Hamiltonian
\begin{equation}
    H_{tun}=\sum_{s,p}\sum_{\sigma v} V_{s}c^\dagger_{sp\sigma v}d_{\sigma v} + h.c
\end{equation}
where $c^\dagger_{sp\sigma v}$ is the creation operator of an electron on side $s$ with momentum $p$, spin $\sigma$ and valley quantum number $v$, and it is assumed that the tunneling process conserves spin and valley degrees of freedom. The electrons in   the leads are described by the non-interacting Hamiltonian
\begin{equation}
    H_{res}=\sum_{s,p}\sum_{\sigma,v} \epsilon_{sp}c^\dagger_{sp\sigma v}c_{sp\sigma v}
\end{equation}
We assume equal tunneling to the left and right reservoirs, $\Gamma_{L/R} =\Gamma= \pi \rho V_{L/R}$, and equal and constant density of states $\rho$ in the two leads, whose bandwidth  is defined as the unit of energy in the calculations.

The numerical calculations required for obtaining the conductance, are performed
using a density matrix numerical renormalization group
(DM-NRG) procedure \cite{toth2008density}
implemented in the Budapest Flexible DM-NRG code
   \cite{legeza2008manual}. 

 We keep 3000 states at every iteration with discretization constant $\Lambda=2$. 
Conductance is then calculated as
\begin{equation}
    G=\frac{e^2}{h}\Gamma\sum_{\sigma,v}\int d\omega \pi A_{\sigma,v} (\omega) \left(- \frac{\partial f(\omega)}{\partial \omega}\right)
\end{equation}
where $f(\omega)$ is the Fermi function and $A_{\sigma,v}$ is the spectral function associated with level $\{\sigma,v\}$.
The differential conductance is then approximated using the equilibrium spectral function 
\begin{equation}
\frac{dI}{dV}=\frac{e^2}{h}\frac{\Gamma}{2}\sum_{\sigma,v}\int d\omega \pi A_{\sigma,v} (\omega)\frac{\partial\left[ -f(\omega+V/2)-f(\omega-V/2)\right]}{\partial\omega}  
\label{didv}
\end{equation}

\begin{figure}
	\includegraphics{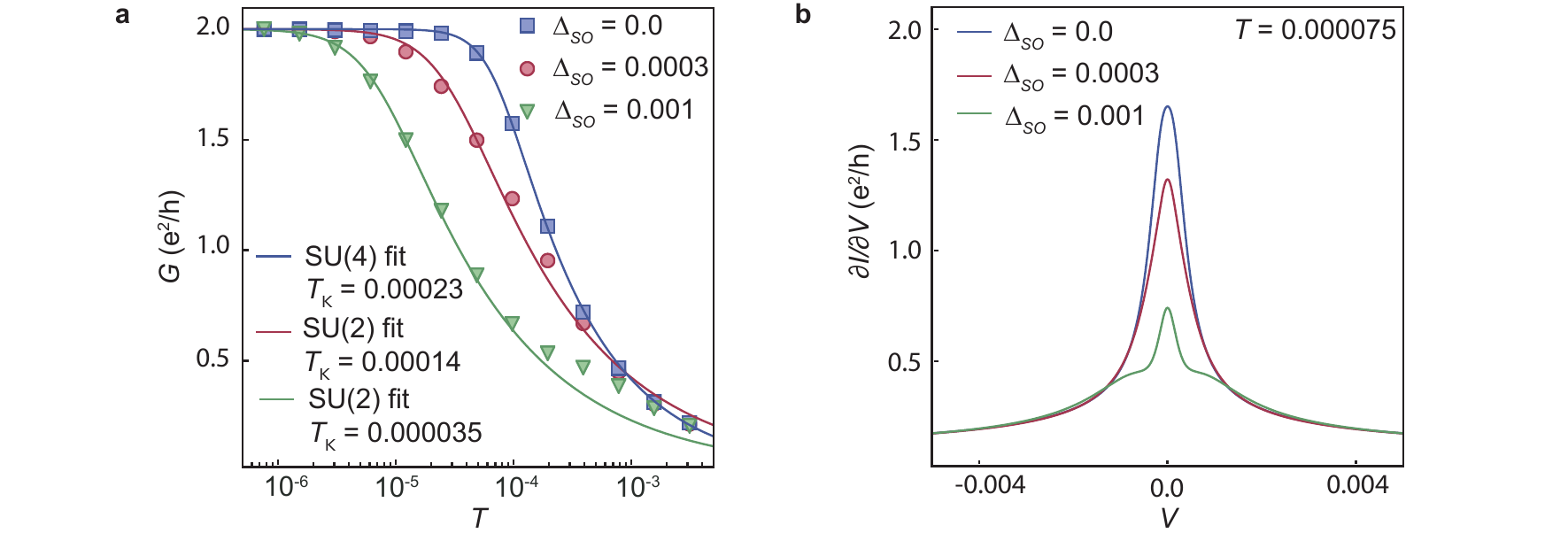}
	\caption{SU(4) to SU(2) Kondo effect for $N=1\,\text{h}$. \textbf{a}, Conductance as a function of temperature shows universal scaling laws in the Kondo regime for various SO interactions strengths ($\Delta_{SO}=0,0.0003,0.001$ for SU(4) - blue, SU(2) - red and SU(2) - green, scaling properties). A system with a $\Delta_{SO}$ that is slightly larger than $T_k$ can already appear to scale closer to SU(2) universal scaling (red points) \textbf{b}, Differential Conductance as a function of bias voltage, calculated using the approximation Eq.~\ref{didv} shows that splitting only occurs for larger SO couplings. Even parameters shown in (a) to exhibit SU(2) scaling (red curve) can still have a single zero-bias peak without side peaks.}
	\label{figs1}
\end{figure}

For single occupation $N=1$ (Fig.~\ref{figs1}) we use $\Gamma=0.04$ and $\epsilon_d=-0.4$, where the Coulomb repulsion is taken infinite, and calculate conductance and differential conductance for three values of spin-orbit gap $\Delta_{so}=0, 0.0003, 0.001$. The red curve in Figs.~\ref{figs1}a,b agrees with the experimental situation, where the degeneracy is lifted such that scaling follows an SU(2) Kondo scaling while, at the same time, the splitting of the zero-bias peak is not yet resolved.

\begin{figure}
	\includegraphics{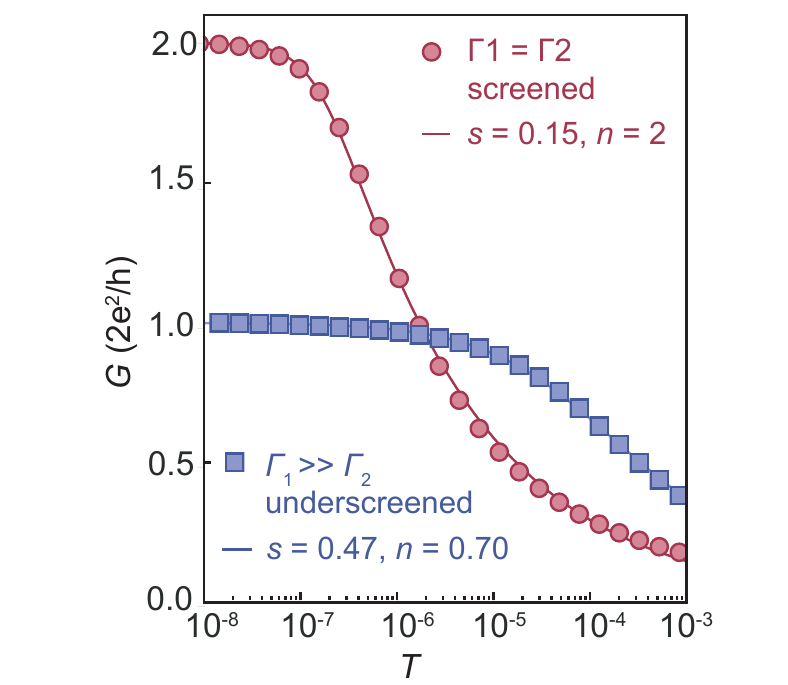}
	\caption{Fitting parameters for $N=2\,\text{h}$. Conductance as a function of temperature in the spin-triplet grounds state regime for various two scenarios: fully screened (red) and underscreened (blue). A purely emirical scaling law (Eq.~1) was used without constraint of n and s.}
	\label{figs2}
\end{figure}

For double occupation $N=2$ (Fig.~\ref{figs2}) we use a Coulomb repulsion of $U=1$ and the level resides at $\epsilon=-1.5$. We use a ferromagnetic interaction $J=-0.1$ to force the system into the spin triplet state forming an $S=1$ effective model. Then, two scenarios are investigated: (a) The underscreened scenario (blue points in Fig.~\ref{fig2}), where just one channel participates in the screening $\Gamma_{k}=0.1,\Gamma_{k'}=0.01$. The k' channel Kondo temperature is below the lowest temperature depicted in the plot. (b) The fully screened scenario (red points in Fig.~\ref{figs2}), where both channels participate equally in the screening $\Gamma_{k}=\Gamma_{k'}=0.04$. This is characterized by a different scaling behavior than the underscreened case. The two scenarios are fitted with the empirical curve given by Eq.~(1) in the main text. This empirical form was also studied for various $S>1/2$ Kondo models for the underscreened case \cite{parks2010mechanical} with an additional fitting parameter that will not be utilized here, so the parameters we get might deviate slightly from that result. The screened case, which is expected to scale according to Fermi liquid quadratic temperature dependence with scaling parameter $n=2$, was studied for S=1 with similar results\cite{blesio2019fully}. The empricial fitting parameters are given in the legend to Fig.~\ref{figs2}. These numerical fits were used in Fig.~3 in the main text to fit the experimental data.

\end{document}